\documentclass[12pt]{article}

\usepackage{graphicx}
\usepackage{epsfig}
\usepackage{amsfonts}
\usepackage{amssymb}
\usepackage{amsmath,amssymb}

\newcommand{\ee}{\end{equation}}
\newcommand{\eea}{\end{eqnarray}}
\newcommand{\be}{\begin{equation}}
\newcommand{\bea}{\begin{eqnarray}}

\begin{document}

\title{  Einstein--Maxwell--Anti-de-Sitter spinning solitons}

\author{
{\large Carlos Herdeiro}\footnote{herdeiro@ua.pt} \
and
{\large Eugen Radu}\footnote{eugenradu@ua.pt}
\\ 
\\
{\small Departamento de F\'\i sica da Universidade de Aveiro and} \\
{\small  Center for Research and Development in Mathematics and Applications (CIDMA)} \\ 
{\small Campus de Santiago, 3810-183 Aveiro, Portugal}
}
\date{February 2016}
\maketitle

\begin{abstract}   
Electrostatics on global Anti-de-Sitter ($AdS$) spacetime is sharply different from that on global Minkowski spacetime. It admits a multipolar expansion with everywhere regular, finite energy solutions, for every multipole moment except the monopole~\cite{Herdeiro:2015vaa}. A similar statement holds for global $AdS$ magnetostatics. We show that everywhere regular, finite energy, electric plus magnetic fields exist on $AdS$ in three distinct classes: $(I)$ with non-vanishing \textit{total} angular momentum $J$; $(II)$ with vanishing $J$ but non-zero angular momentum \textit{density}, $T^t_\varphi$; $(III)$ with vanishing $J$ $and$ $T^t_\varphi$. 
Considering backreaction, these configurations remain everywhere smooth and finite energy, and 
 we find, for example, 
 Einstein--Maxwell--$AdS$ solitons that are  \textit{globally} -- Type I -- or \textit{locally} (but not globally) -- Type II -- spinning. This backreaction is considered first perturbatively, using analytical methods and then non-perturbatively, by constructing numerical solutions of the fully non-linear Einstein--Maxwell--$AdS$ system. The variation of the energy and total angular momentum with the boundary data is explicitly exhibited for one example of a spinning soliton.
\end{abstract}

\section{Introduction} 
In a recent letter~\cite{Herdeiro:2015vaa} we have shown that electrostatics on global $AdS$ presents two important differences from standard electrostatics on Minkowski spacetime. Firstly, all multipole moments (except for the monopole) are everywhere regular and finite energy. Secondly, all multipole moments decay with the same inverse power of the areal radius, $1/r$, as spatial infinity is approached. The first observation 
suggests the existence of regular, self-gravitating, asymptotically $AdS$ Einstein--Maxwell solitons, obtained as the non-linear backreacting versions of these regular electric multipoles; the second observation renders inapplicable Lichnerowicz-type no-soliton theorems~\cite{Boucher:1983cv,Shiromizu:2012hb}.  Such Einstein--Maxwell--$AdS$ \textit{static} solitons indeed exist, and examples were constructed perturbatively in~\cite{Herdeiro:2015vaa} and nonperturbatively in~\cite{Costa:2015gol}. 

Typically, static gravitating solitons allow for spinning generalizations; however, see~\cite{VanderBij:2001nm,Radu:2008pp}.  
Thus, in this letter, we address the existence of Einstein--Maxwell--$AdS$ spinning solitons. A simple reasoning shows the way forward.

Given the aformentioned results for electrostatics on global $AdS$, electromagnetic duality implies that magnetostatics on global $AdS$ also presents everywhere regular, finite energy solutions. We shall explicitly verify it is so. Moreover, at test field level, the superposition principle allows electric plus magnetic configurations which, again, are everywhere regular and with finite energy. The latter have, in general, a non-zero Poynting vector, $i.e.$ a non-zero angular momentum density. As we show below, however, the existence of a local Poynting vector does not imply a non-zero global angular momentum; that only happens for the particular case when ``next neighbour" electric and magnetic multipoles occur in the superposition. Then we consider the backreaction of these electromagnetic fields with non-vanishing total angular momentum, and construct, both perturbatively (analytically) and non-perturbatively (numerically) the corresponding \textit{spinning} Einstein--Maxwell--$AdS$ solitons.

\section{The model: Einstein--Maxwell--$AdS$ theory}
\label{sec_2}
Following~\cite{Herdeiro:2015vaa}, we shall be addressing Einstein--Maxwell theory in the presence of
a negative cosmological constant (hereafter dubbed \textit{Einstein-Maxwell-AdS gravity}), described by the action:
\begin{eqnarray}
\mathcal{S} =\int d^4 x\sqrt{-g}\left\{\frac{1}{16\pi G}\left(R-2\Lambda\right)
 -\frac{1}{4}F_{\mu \nu}F^{\mu\nu}\right\} \ ,
 \label{EMAdS}
\end{eqnarray}
where $F=d{\cal A}$ is the $U(1)$ Maxwell field strength, $\Lambda\equiv -3/L^2<0$ 
is the negative cosmological constant and $L$ is the $AdS$ ``radius". Varying the action one obtains the Maxwell equations
\begin{eqnarray}
d\star F=0 \ ,
\label{me}
\end{eqnarray}
and the Einstein equations
 \begin{eqnarray} 
 R_{\mu\nu}-\frac{1}{2}R g_{\mu\nu}+\Lambda g_{\mu\nu}=8\pi G~ T_{\mu\nu}  \ ,
\label{eineq}
\end{eqnarray} 
where $T_{\mu\nu}$ is the electromagnetic energy-momentum tensor 
\begin{eqnarray}
T_{\mu\nu}=F_{\mu \alpha}F_{\nu\beta}g^{\alpha \beta}-\frac{1}{4}g_{\mu\nu}F^2\ .
\label{emtensor}
\end{eqnarray}
The background of our model is the (maximally symmetric) $AdS$ spacetime, with $F=0$. 
In global coordinates it takes the form
\begin{eqnarray}
\label{AdS}
ds^2=-N(r)dt^2+\frac{dr^2}{N(r)}+r^2(d \theta^2+\sin^2\theta d\varphi^2) \ 
,~~{\rm where}~~N(r)\equiv 1 +\frac{r^2}{L^2} \ .
\end{eqnarray}

\section{Test fields: electro-magnetostatics on $AdS$}
\label{sec_3}
We start by considering linear Maxwell perturbations around  an empty $AdS$ background. Thus we solve the (test) Maxwell equations (\ref{me}) on the geometry (\ref{AdS}). For time-independent, axially symmetric Maxwell fields, a suitable gauge potential ansatz reads
\begin{eqnarray}
{\cal A}\equiv {\cal A}_\mu dx^\mu=V(r,\theta) dt+A(r,\theta)  d\varphi \ . 
\label{ep}
\end{eqnarray}

\subsection{Static solutions} 
Let us start with the simplest case: either a purely electric or a purely magnetic field, but not both simultaneously. 
Then the Poynting vector vanishes and the solutions carry no angular momentum.

\subsubsection{Electrostatics on global $AdS$} 
This case has been considered in \cite{Herdeiro:2015vaa}.
Here we review its basic properties.
The axisymmetric electric potential in~\eqref{ep} 
can be expressed as a multipolar expansion 
\begin{eqnarray}
{\cal A}_t\equiv V(r,\theta)=\sum_{\ell=0}^\infty c_E^{(\ell)} V_{\ell}(r,\theta) \ , \qquad \  V_{\ell}(r,\theta)\equiv R_\ell(r) \mathcal{P}_\ell(\cos \theta)\ , 
\label{epe}
\end{eqnarray}
where 
 $\mathcal{P}_\ell$ is a Legendre polynomial of degree $\ell$ (with $\ell\in \mathbb{N}_0$ defining the multipolar structure) and $c_E^{(\ell)}$ are arbitrary constants.
Then Maxwell's equations reduce to the radial equation
\begin{eqnarray}
\label{eq}
\frac{d}{dr} \left(r^2\frac{dR_\ell(r)}{dr}\right)=\frac{\ell(\ell+1)}{N(r)}R_\ell\ .
\label{radial}
\end{eqnarray}
An everywhere regular solution of this equation  is found for $\ell\geqslant 1$, with
\begin{eqnarray}
R_\ell(r)=
\displaystyle{\frac{\Gamma(\frac{1+\ell}{2})\Gamma(\frac{3+\ell}{2})}{\sqrt{\pi}\Gamma(\frac{3}{2}+\ell)}
 \frac{r^{\ell}}{L^\ell}~{}_2F_1\left(\frac{1+\ell}{2} , \frac{ \ell}{2}   ;  \frac{3}{2}+\ell  ;  - \frac{r^2}{L^2}\right)} \ ,
\end{eqnarray} 
expressed in terms of the hypergeometric function ${}_2F_1$ and normalized such that $R_\ell(r)\to 1$ asymptotically.

At the origin, the $AdS$ regular multipoles approach the behaviour of the  Minkowski multipoles that are regular therein: 
 \begin{eqnarray}
R_\ell(r)= 
\frac{ \Gamma\left(\frac{1+\ell}{2}\right) \Gamma\left(\frac{3+\ell}{2}\right)}{\sqrt{\pi} 
\Gamma\left(\frac{3}{2}+\ell\right)} 
\left(\frac{r}{L}\right)^\ell
+\dots \ .
\end{eqnarray} 
Asymptotically, however, the regular $AdS$ multipoles are very different from the Minkowski multipoles which are regular at infinity. As $r\rightarrow \infty$, the solutions become 
\begin{eqnarray}
\label{as1}
R_\ell(r)=1-\frac{2 \Gamma\left(\frac{1+\ell}{2}\right) \Gamma\left(\frac{3+\ell}{2}\right)}{ \Gamma\left(1+\frac{\ell}{2}\right) \Gamma\left(\frac{\ell}{2}\right)}
\frac{L}{r}+\dots\ . 
%
\end{eqnarray} 
thus, all multipoles fall-off with the same $1/r$ power, where $r$ is the areal radius, $cf.$ eq. (\ref{AdS}).

 The total energy of each regular electric multipole can actually be expressed as a surface integral. Noticing that 
\begin{eqnarray}
E_e = -\pi \lim_{r\to \infty}\int_0^\pi  r^2 \sin \theta {\cal A}_t F^{rt} d\theta \ ,
\end{eqnarray} 
we obtain, for a given multipole $\ell$,  
\begin{equation}
\label{energy-el}
E_{e}^{(\ell)}=\frac{4\pi}{2\ell+1}\frac{ \Gamma(\frac{1+\ell}{2}) \Gamma(\frac{3+\ell}{2})}{ \Gamma(1+\frac{\ell}{2}) \Gamma(\frac{\ell}{2})}  L \ .
\end{equation}

\subsubsection{Magnetostatics on global $AdS$} 
Due to the electric-magnetic duality of Maxwell's theory, which leaves invariant~\eqref{EMAdS}, the configurations  of the previous subsection possess an equivalent magnetic picture in terms of the potential $A(r,\theta)$ in~\eqref{ep} 
(and a vanishing $V(r,\theta)$).
Thus, for each electric $\ell$-multipole (\ref{epe}), 
one finds a dual magnetic $\ell$-multipole solution of Maxwell's equations, described by
\begin{eqnarray}
  A_\ell (r,\theta)=S_{\ell}(r)U_{\ell}( \theta)\ ,
\end{eqnarray}
where
 \begin{eqnarray}
S_{\ell}(r)=r^2 \frac{dR_{\ell}(r)}{dr} \ ,~~\qquad U_{\ell}( \theta)=\sin \theta \frac{d  \mathcal{P}_\ell(\cos \theta)}{d\theta} \ .
\end{eqnarray}
Observe the absence of the Dirac string on the symmetry axis. The general, \textit{everywhere regular}, magnetic potential in~\eqref{ep} is a superposition of all these $\ell\geqslant 1$ multipoles  (with $c_M^{(\ell)}$ arbitrary constants)
\begin{equation}
{\cal A_\varphi}\equiv A(r,\theta)=\sum_{\ell=1}^\infty c_M^{(\ell)} A_\ell (r,\theta) \ .
\label{ageneral}
\end{equation}

The explicit form of the functions $S_{\ell}(r)$ looks more complicated than in the electric case:
 \begin{eqnarray}
S_{\ell}(r)&=&L \frac{\ell \Gamma(\frac{\ell}{2}+1) \Gamma(\frac{\ell}{2} )}{2\sqrt{\pi} \Gamma( \ell+\frac{3}{2}) }
\left( \frac{r}{L}\right)^{\ell+1}
 \bigg [
 {}_2 F_1\left( \frac{1+\ell}{2},   \ell  ;  \frac{3}{2}+\ell  ;  - \frac{r^2}{L^2}\right)   
\\
\nonumber
&&{~~~~~~~~~~~~~~~~~~~~~~~~~~~~}-\frac{\ell+1}{2\ell+3} 
{~}_2 F_1\left( \frac{3+\ell}{2},   \frac{2+\ell}{2} ;  \frac{5}{2}+\ell  ;  - \frac{r^2}{L^2}\right)  \frac{r^2}{L^2} 
 \bigg ].
\end{eqnarray}
Here, the solution is normalized such that $S_{\ell}(r)\to L$ as $r\to \infty$ 
and the factor of $L$ is introduced for dimensional reasons.

As $r\to 0$, the radial part of the  magnetic potential behaves as
 \begin{eqnarray}
S_\ell(r)=  L\left(\frac{r}{L}\right)^{\ell+1}   \frac{\Gamma(\frac{\ell }{2}+1)^2}{\sqrt{\pi} \Gamma(\frac{3}{2}+\ell)}  +\dots \ ,
\end{eqnarray} 
while its far field expression is
 \begin{eqnarray}
S_\ell(r)=  L\left \{1- \frac{2L}{r} \left[\frac{\Gamma(\frac{\ell }{2}+1)}{ \Gamma(\frac{\ell+1}{2})} \right]^2
\right\}+\dots \ .
\label{as2}
\end{eqnarray}

Again, the total energy can be expressed as a surface integral, by noticing that 
\begin{eqnarray}
E_{m} = -\pi \lim_{r\to \infty}\int_0^\pi  r^2 \sin \theta {\cal A}_\varphi F^{r \varphi} d\theta \ .
\end{eqnarray} 
 Then, for a given multipole $\ell$, 
we obtain\footnote{Observe that the energies of the regular electric and magnetic $\ell$-multipoles, eqs.~\eqref{energy-el} and \eqref{energy-mag}, respectively, are different. This does not contradict the fact that electric-magnetic
duality implies the energy density (and total energy) of dual $\ell$-modes must match. The difference arises due to the  chosen normalization of the radial function $S_{\ell}(r)$, which would be different in case the magnetic modes were computed directly from the duality transformation.}
\begin{equation}
\label{energy-mag}
E^{(\ell)}_m=\frac{4\pi \ell(\ell+1)}{2\ell+1}\left[\frac{ \Gamma(\frac{1+\ell}{2})  }{ \Gamma( \frac{\ell+1}{2}) } \right]^2  L \ .
\end{equation}

\subsection{Stationary solutions}
We now turn to generic axially symmetric  configurations, consisting in the superposition of an electric potential
\textit{plus} a magnetic potential. Considering all possible {\it regular} electric and magnetic modes, the general expression for the gauge potential is:
 \begin{eqnarray}
 \label{sol}
{\cal A}=\sum_{\ell \geq 1}c_E^{(\ell)} R_\ell(r)\mathcal{P}_\ell(\cos \theta)dt
+\sum_{p \geq 1}c_M^{(p)} S_{p}(r)U_{p}( \theta)d\varphi \ .
\end{eqnarray}

\begin{figure}[h!]
\begin{center}
\includegraphics[width=0.6\textwidth]{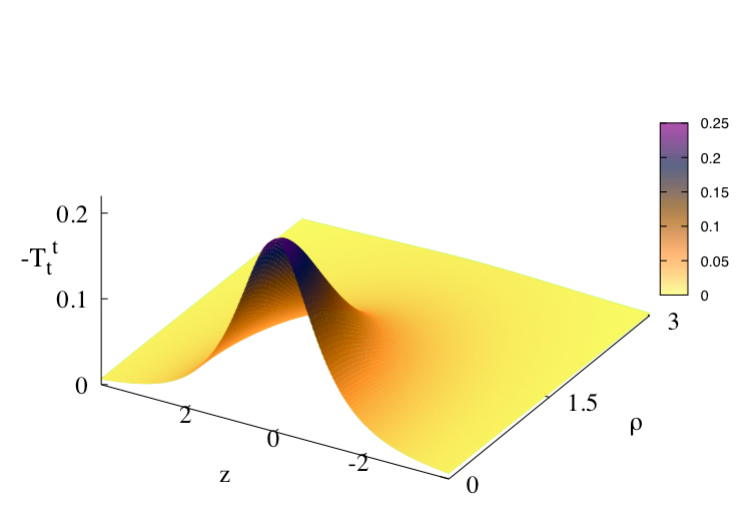}
\includegraphics[width=0.6\textwidth]{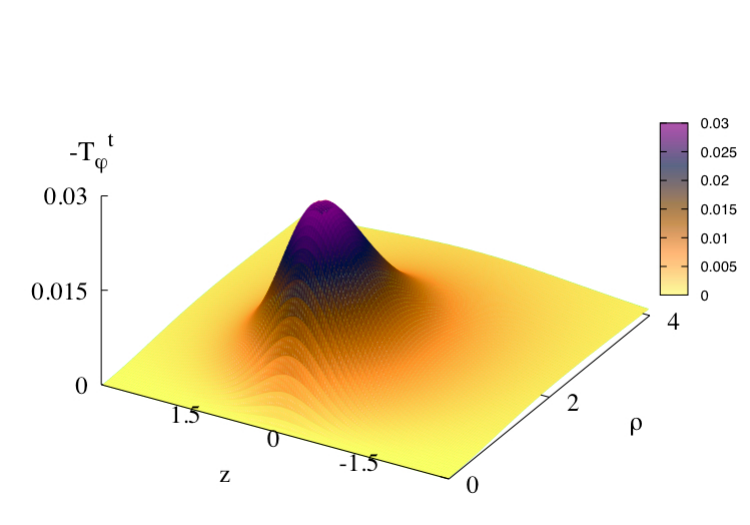}
\caption{The energy density ($-T_t^t$) and angular momentum density ($T_\varphi^t$)
(with a minus sign for a better visualization)
are shown as functions of ``cylindrical" coordinates $\rho\equiv r\sin\theta$ and $z\equiv r\cos\theta$,  for an electric plus magnetic Maxwell field consisting on the superposition of a magnetic $p=1$ and an electric $\ell=2$ multipole.}
\label{e2m1}
\end{center}
\end{figure}

The total energy of the electro-magnetic configurations is obtained by adding the energy of the corresponding electric and magnetic modes 
 \begin{eqnarray}
\label{M1}
E=  \sum_{\ell \geq 1} (c_E^{(\ell)})^2 E_e^{(\ell)}+ \sum_{p \geq 1} (c_M^{(p)})^2 E_m^{(p)}\ ,
\end{eqnarray}
in accordance with the superposition principle, where  $E_e^{(\ell)}$ and $ E_m^{(p)}$ are
given by  (\ref{energy-el}) and (\ref{energy-mag})
respectively. But turning on simultaneously the electric and magnetic fields can also yield a non-trivial Poynting vector and consequently angular momentum. To check whether this happens or not, we consider the angular momentum density of the above solution
 \begin{eqnarray}
T_\varphi^t=F_{r \varphi}F^{rt}+F_{\theta \varphi}F^{\theta t}\ ,
\end{eqnarray}
and the corresponding total angular momentum:
  \begin{eqnarray}
J=2\pi \int_0^\infty   r^2 dr \int_0^\pi \sin \theta T_\varphi^t d\theta \ .
\end{eqnarray}
By using Maxwell's equations it follows that $J$ is given by the boundary integral
\begin{eqnarray}
J=2\pi \lim_{r\to \infty}\int_0^\pi   r^2 \sin \theta {\cal A}_\varphi F^{rt}d\theta \ .
\end{eqnarray}
This general expression can be evaluated by using the far field expressions~\eqref{as1} and~\eqref{as2},
together with the properties  of the Legendre polynomials.
One finds
 \begin{equation}
\label{J}
J=2\pi L^2 \sum_{p \geq 1}c_E^{(\ell)} c_M^{(p)}
\left\{
 \frac{8(\ell+2)\left[\Gamma(\frac{\ell+3}{2})\right]^2}
{(2\ell+1)(2\ell+3)\Gamma(\frac{\ell }{2}+1)\Gamma\left(\frac{\ell }{2} \right) }\delta_{p,\ell+1}
-\frac{8(\ell-1)\Gamma(\frac{\ell+1}{2}) \Gamma(\frac{\ell+3}{2})}{(4\ell^2-1)\left[\Gamma(\frac{\ell }{2})\right]^2}
\delta_{p,\ell-1} 
\right\} \ .
\end{equation}
Interestingly,  the total angular momentum vanishes unless there are ``next neighbours" electric and magnetic multipoles in the superposition. One example of this type of configuration, with non-zero total angular momentum, 
which we call \textit{Type I} (electric plus magnetic regular Maxwell field), is given in Fig.~\ref{e2m1}, where we exhibit the energy density ($-T^t_t$) and angular momentum density ($T^t_\varphi$) 
for $\ell=2$ and $p=1$. 

\begin{figure}[h!]
\begin{center}
\includegraphics[width=0.6\textwidth]{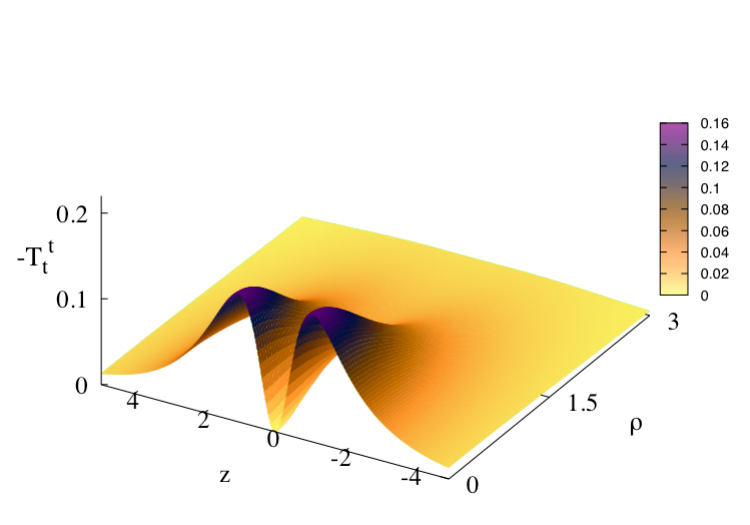}
\includegraphics[width=0.6\textwidth]{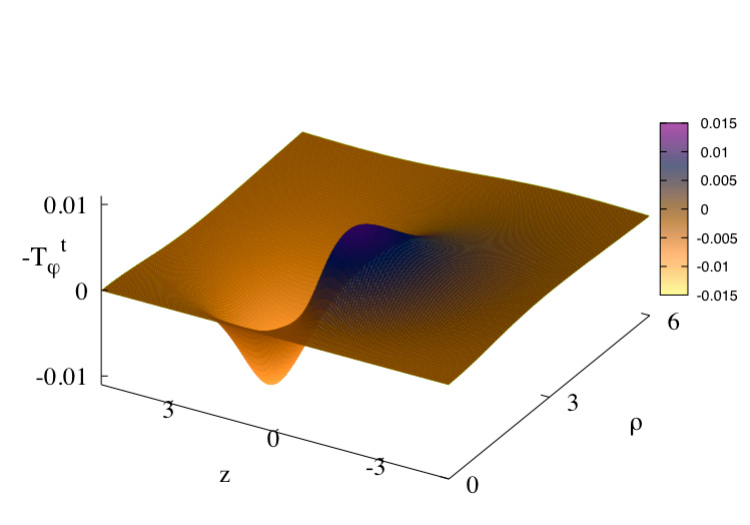}
\caption{Same as in Fig. \ref{e2m1}, but now the electric plus magnetic Maxwell field consists on the superposition of a magnetic $p=2$ and an electric $\ell=4$ multipole.}
\label{e4m2}
\end{center}
\end{figure}

For the cases where there are both electric and magnetic multipoles but not ``next neighbours", the total angular momentum vanishes; the angular momentum \textit{density}, however, in general does not. 
One example of this type of configurations, with zero total angular momentum but non-vanishing angular momentum density, which we call \textit{Type II} (electric plus magnetic regular Maxwell field), is given in Fig.~\ref{e4m2}, 
where it can be seen that the angular momentum density is odd under the $\mathbb{Z}_2$ transformation $z\rightarrow -z$. 
This explains why the total angular momentum vanishes. 
Observe also that the energy density exhibits two distinct lumps, 
each corresponding to a different sign of the angular momentum density, whereas in the case of Fig.~\ref{e2m1} 
there is a single lump.\footnote{It is interesting to notice the 
existence of similar solutions in a flat space Yang-Mills--Higgs theory 
\cite{Paturyan:2004ps,Kleihaus:2005fs}.
According to the terminology above,
the dyons of that model are $Type~II$ solutions ($i.e.$ they can spin only locally
but not globally),
while the composite configurations with a vanishing net magnetic charge
are $Type~I$ solutions.
}

Finally, for configurations with $p=\ell\geq 1$, the angular momentum density vanishes identically, 
$T_\varphi^t=0$. This type of configurations 
(which is akin to the
well-known spherically symmetric Maxwell dyon), are called \textit{Type III} (electric plus magnetic regular Maxwell field). These configurations, however, are duality trivial, 
in the sense that either the electric or the magnetic component 
can be eliminated by a duality transformation.

\section{Backreacting solutions: Einstein--Maxwell--$AdS$ solitons}
The existence of everywhere regular, finite energy Maxwell fields on global $AdS$, as described in the previous section, some of which have non-zero total angular momentum, suggests the existence of fully non-linear Einstein--Maxwell--$AdS$ spinning solitons, as backreacting non-linear versions of the test field solutions. We shall now consider such backreaction starting with an analytic perturbative approach and then constructing them numerically, at fully non-linear level.

\subsection{A perturbative analytic approach}
Following the approach in~\cite{Herdeiro:2015vaa} for the static case, we start testing the effects of backreaction by constructing a perturbative solution to the Einstein--Maxwell system. We consider a perturbative parameter $\alpha$, standing for the ``magnitude"  of \textit{both} electric and magnetic potentials at infinity, thus taken to be equal. 

The perturbative solutions are studied with the following line element, which yields a convenient metric gauge choice:
 \begin{eqnarray}
\label{nex1}
ds^2=-F_1(r,\theta)N(r)dt^2+F_2(r,\theta)\frac{dr^2}{N(r)}+F_3(r,\theta) r^2 
\left[
d\theta^2+\sin^2\theta (d\varphi+W(r,\theta)dt)^2
\right] \ .
\end{eqnarray}
This corresponds to deforming the pure $AdS$ line element~\eqref{AdS} with four functions, $F_1,F_2,F_3, W$, all depending on $r,\theta$ only. 
The gauge potential ansatz, on the other hand, is still taken to be of the form~\eqref{ep}, thus depending on the two functions $V,A$, 
of $r,\theta$. 
These gauge potentials, however, are now expanded in a power series in $\alpha$. 
Up to $\mathcal{O}(\alpha^3)$, the expansion reads [$\mathcal{O}(\alpha^2)$ terms vanish]:
 \begin{eqnarray}
\label{nex2}
V(r,\theta) =\alpha V^{(0)}(r,\theta) +  \alpha^3 V^{(3)}(r,\theta)+\dots\ ,~~
A(r,\theta) =\alpha A^{(0)}(r,\theta) +  \alpha^3 A^{(3)}(r,\theta)+\dots\ ,~~
\end{eqnarray}
where $V^{(0)}(r,\theta)$ and $A^{(0)}(r,\theta)$ are the (general linear combination of) test Maxwell fields on $AdS$ studied in the previous section, given by~\eqref{epe} and~\eqref{ageneral}, respectively. The backreaction of these Maxwell fields on the geometry is taken into account by considering a power series expansion in $\alpha$ of the metric functions, of the form:
%
 \begin{eqnarray}
\label{nex3}
 F_i(r,\theta) =1+\alpha^2 F_{i2}(r,\theta)+ \alpha^4 F_{i4}(r,\theta)+\dots,~~
W(r,\theta) = \alpha^2 W_{2}(r,\theta)+ \alpha^4 W_{4}(r,\theta)+\dots,~~
\end{eqnarray}
and solving the coupled Einstein--Maxwell equations order by order. 
The test field solution ($\mathcal{O}(\alpha)$ in this setup) supplies
 the boundary condition for ${\cal A}$ at the $AdS$ boundary.

To illustrate this perturbative procedure, let us briefly discuss it for the case one fixes, at the $AdS$ boundary,  the superposition of a $p=1$ magnetic mode and  an $\ell=2$ electric mode. Thus the $\mathcal{O}(\alpha)$ data is
 \begin{eqnarray}
A^{(0)}(r,\theta) &=& -c_m L \left[1-\frac{L}{r}\arctan \left(\frac{r}{L}\right) \right]\sin^2\theta\ ,
\\
\nonumber
V^{(0)}(r,\theta) &=&c_e 
\left[ 
1+\frac{3L^2}{2r^2}-\frac{3L}{2r}\left(1+\frac{L^2}{r^2}\right)\arctan\left(\frac{r}{L}\right)
\right]\left[ \frac{3}{2}\cos^2\theta- \frac{1}{2}\right] \ ,
\end{eqnarray}
where $c_e$, $c_m$ are two arbitrary constants.

Then,
 the perturbed solution is constructed by taking the metric perturbations as an angular expansion in Legendre functions with coefficients given by radial functions,  the  expression to lowest order being
 \begin{eqnarray}
\nonumber
F_{i2}(r,\theta)&=&a_{i}(r)+\mathcal{P}_2(\cos\theta)b_{i}(r)+\mathcal{P}_4(\cos\theta)c_{i}(r)\ , 
\label{nex31} 
\\
\nonumber
W_2(r,\theta)&=&U_0(r)+\mathcal{P}_2(\cos\theta)U_{2}(r)\ ,
\end{eqnarray}
with $i=1,2,3$, and the gauge potential functions expanded in a similar way
 \begin{eqnarray}
\nonumber 
A^{(3)}(r,\theta)=\sum_{k=0}^3 \mathcal{P}_{2k}(\cos\theta)g_{2k}(r)\ ,~~
V^{(3)}(r,\theta)=\sum_{k=0}^3 \mathcal{P}_{2k}(\cos\theta)h_{2k}(r) \ . 
\end{eqnarray}
In solving the Einstein equations, one uses a residual gauge freedom to set the radial function $a_3=0$.
The remaining radial functions are found in closed form by solving the Einstein--Maxwell equations order by order in $\alpha$.
When doing so, one requires preservation of the $AdS$ asymptotics;
for the gauge field we impose $A \to -c_m L \sin^2\theta$ and  $V\to c_t ( \frac{3}{2}\cos^2\theta- \frac{1}{2})$ as $r\to \infty$.

We have solved  for $F_i$ to $\mathcal{O}(\alpha^2)$ and for $A$, $V$, $W$ to $\mathcal{O}(\alpha^4)$.
The explicit form of the solutions is very long and not enlightening, \textit{per se}; thus, we shall not display them here.
The only expressions which take a  simpler form  are for the functions that enter $W_{2}(r,\theta)$, and read
(here we set $4 \pi G=1$)
 \begin{eqnarray}
\nonumber
 U_0(r)&=&\frac{c_e c_m }{32L}
\left\{
2\pi^2\left(1+\frac{L^2}{r^2}+\frac{3}{2}\frac{L^4}{r^4}\right)
-\frac{8L^2}{r^2}\left(1+\frac{3L^2}{r^2}\right) \right.
\\
\nonumber
&&
\left.
-\left(1+\frac{ L^2}{r^2}\right)\arctan\left(\frac{r}{L}\right)\left\{\frac{L}{r}\left[16+3(\pi^2-16)\frac{L^2}{r^2}\right] 
+8\left(1+\frac{ 3L^4}{r^4}\right) \arctan\left(\frac{r}{L}\right)
\right\}\right\},
\\
\nonumber
 U_2(r)&=&\frac{c_e c_m }{32L}
\left\{
2\pi^2\left(-1+\frac{5L^2}{r^2}+\frac{15}{2}\frac{L^4}{r^4}\right)
+\frac{8L^2}{r^2}\left(1-\frac{3L^2}{r^2}\right)\right.
\\
\nonumber
&&
\left.+\left(1+\frac{ L^2}{r^2}\right)\arctan\left(\frac{r}{L}\right)\left\{\frac{L}{r}\left[16+3(16-5\pi^2)\frac{L^2}{r^2}\right] 
+8\left(1-\frac{ 6L^2}{r^2}-\frac{ 3L^4}{r^4}\right) \arctan\left(\frac{r}{L}\right)
\right\}\right\}.
\end{eqnarray}
We have checked that 
the backreacted metric is smooth and no evidence of pathologies has been found.
Moreover,
we have verified that
the spacetime is asymptotically $AdS$,
according to the definition in  
\cite{Ashtekar:1999jx}.
The gauge potentials are also smooth.
The total angular momentum has a sufficiently compact expression, which reads:
 \begin{equation}
 J=- c_e c_m \pi^2 \alpha^2 
L^2
\left\{\frac{2}{5}  
+\alpha^2  
\left[
c_m^2 \frac{(32000+48 \pi^2-405 \pi^4)}{40320 }
+c_e^2 \frac{ (1024000+702144\pi^2+5265\pi^4)}{2150400}
\right]\right\} .
\end{equation}

\subsection{The fully non-linear numerical approach}
 In the absence of analytic methods to tackle the fully non-linear Einstein--Maxwell 
spinning solitons described in the previous sections,
 we shall resort to numerical methods.\footnote{Unfortunately, 
the powerful analytical techniques used to integrate the $\Lambda=0$  
axially symmetric  Einstein--Maxwell system cannot be extended to the $AdS$ case 
\cite{Charmousis:2006fx,Astorino:2012zm}.}
 
\subsubsection{Framework}
Non-perturbative solutions will be constructed by employing the Einstein--De Turck (EDT) approach, proposed in~\cite{Headrick:2009pv,Adam:2011dn}. This approach has become, in recent years, a standard tool in the numerical treatment of stationary problems in general relativity,
and has the advantage of not fixing $apriori$ 
a metric gauge, 
yielding at the same time elliptic equations
(see~\cite{Wiseman:2011by,Dias:2015nua} for reviews).   
Then, instead of (\ref{eineq}), one solves the so called EDT equations
\begin{eqnarray}
\label{EDT}
R_{\mu\nu}-\nabla_{(\mu}\xi_{\nu)}=\Lambda g_{\mu\nu}+8 \pi G \left(T_{\mu\nu}-\frac{1}{2}T  g_{\mu\nu}\right) \ .
\end{eqnarray}
Here, $\xi^\mu$ is a vector defined as 
$
\xi^\mu\equiv g^{\nu\rho}(\Gamma_{\nu\rho}^\mu-\bar \Gamma_{\nu\rho}^\mu)\ ,
$
where 
$\Gamma_{\nu\rho}^\mu$ is the Levi-Civita connection associated to the
spacetime metric $g$ that one wants to determine, and a reference metric $\bar g$ is introduced, 
($\bar \Gamma_{\nu\rho}^\mu$ being the corresponding Levi-Civita connection).
Solutions to (\ref{EDT}) solve the Einstein equations
iff $\xi^\mu \equiv 0$ everywhere on
${\cal M}$.
To achieve this,
we impose boundary conditions  which are compatible with
$\xi^\mu = 0$
on the boundary of the domain of integration.
Then, this should imply $\xi^\mu \equiv 0$ everywhere,
a condition which is verified from  the numerical output.

In our approach, we use  a metric ansatz
with six functions, $f_1,f_2,f_3,S_1,S_2,W$,
\begin{eqnarray}
\label{metric}
ds^2&=&-f_0(r,\theta)N(r)dt^2+f_1(r,\theta)\frac{dr^2}{N(r)}+S_1(r,\theta)[rd\theta+S_2(r,\theta)dr]^2 \nonumber \\
&&+f_2(r,\theta)r^2\sin^2\theta \left(d\varphi+\frac{W(r,\theta)}{r}dt\right)^2 \ .
\end{eqnarray}
The obvious reference metric is empty $AdS$, described by the line element (\ref{AdS}), which corresponds to take
$
S_1=f_1=f_2=f_0=1,~~S_2=W=0.
$
The Maxwell field Ansatz is still given by (\ref{ep}) in terms of two potentials,
an electric one $V(r,\theta)$ and a magnetic one,  $A(r,\theta)$.

The EDT equations (\ref{EDT}) together with Maxwell's equations (\ref{me})
result in a set of  8 elliptic partial differential equations which are solved numerically
as a boundary value problem.
The boundary conditions are found by constructing an approximate form
of the solutions on the boundary of the domain
of integration compatible with the requirement $\xi^\mu = 0$ and regularity of the solutions.

Starting with  the $U(1)$ potential, one finds that both   $A$ and $V$ vanish at $r=0$; at $\theta=0,\pi$
one imposes Neuman boundary conditions, 
$\partial_\theta V\big|_{\theta=0,\pi}=\partial_\theta A \big|_{\theta=0,\pi}$.
As $r\to \infty,$ the components of the $U(1)$ potential read
\begin{eqnarray}
\nonumber
V=V^{(0)}(\theta)+\frac{V^{(1)}(\theta)}{r}+\dots,~~A =A^{(0)}(\theta)+\frac{A^{(1)}(\theta)}{r}+\dots,
\end{eqnarray}
where $V^{(0)},A^{(0)}$ are imposed as boundary conditions,
\begin{eqnarray}
\label{condinf-U1}
V\big|_{r=\infty}= V^{(0)}(\theta)\ ,~~A\big|_{r=\infty}= A^{(0)}(\theta) \ , 
\end{eqnarray}
 and 
$V^{(1)},A^{(1)} $
result from the numerical output.
 
 The boundary conditions satisfied by the metric functions at the origin read
\begin{eqnarray}
\partial_r f_1\big|_{r=0}=\partial_r f_2\big|_{r=0}=\partial_r f_0\big|_{r=0}
=\partial_r S_1\big|_{r=0}=\partial_r S_2\big|_{r=0}= W\big|_{r=0}=0\ ,
\end{eqnarray} 
whereas the boundary conditions at the symmetry axis are
\begin{eqnarray}
\partial_\theta f_1\big|_{\theta=0,\pi}=\partial_\theta f_2\big|_{\theta=0,\pi}=
\partial_\theta f_0\big|_{\theta=0,\pi}=\partial_\theta S_1\big|_{\theta=0,\pi}=S_2\big|_{\theta=0,\pi}=
 \partial_\theta W\big|_{\theta=0,\pi}=0\ .
\end{eqnarray} 
The far field behaviour of the functions which enter the line element (\ref{metric}) can be constructed in a systematic way.
The expressions for the functions of interest are
\begin{eqnarray}
\label{far-field}
\nonumber
 &&
 f_0=1+\frac{f_{03}(\theta)}{r^3}+\dots, 
 f_2=1+\frac{f_{23}(\theta)}{r^3}+ \dots,~
~
 S_1=1+\frac{s_{13}(\theta)}{r^3}+\dots,~W=\frac{w_{2}(\theta)}{r^2}+\dots,
\end{eqnarray} 
while $f_1$ decays faster than $1/r^3$ and $S_2$ faster than $1/r^4$.
 $f_{03}(\theta)$, $f_{23}(\theta)$, $s_{13}(\theta)$ and $w_{2}(\theta)$
are 
functions fixed by the numerics,  with $f_{03}(\theta)+f_{23}(\theta)+s_{13}(\theta)=0$. 
Thus, at infinity we impose, as boundary conditions for the metric functions,
\begin{eqnarray}
\label{condinf}
 f_0\big|_{r=\infty}= f_1\big|_{r=\infty}=f_2\big|_{{r=\infty}}= S_1\big|_{r=\infty}=1,~S_2\big|_{r=\infty}=W\big|_{r=\infty}=0\ .
\end{eqnarray}

The mass and angular momentum of the solutions are computed by employing the boundary counterterm approach in \cite{Balasubramanian:1999re}, wherein they are the conserved charges associated with Killing symmetries
$\partial_t$, $\partial_\varphi$ of the induced boundary metric,  found for a large value $r=$constant.
 A straightforward computation leads to the following expressions:
\begin{eqnarray}
\nonumber
M=\frac{3}{8 G L^2}
\int_0^\pi [f_{23}(\theta)+s_{13}(\theta)]\sin \theta  d\theta\ ,
\qquad 
J=-\frac{3}{8 G }
\int_0^\pi  \sin^3 \theta w_2(\theta) d\theta \ .
\end{eqnarray}
Note that the same result can be derived by using the Ashtekar-Magnon-Das conformal mass definition
\cite{Ashtekar:1999jx}.
Moreover, an equivalent expression for the angular momentum is found from the Komar integral:
\begin{equation}
\label{j1}
 J=\frac{1}{8 \pi G }\int R_{\varphi}^t \sqrt{-g} dr  d\theta d\varphi =\int T_{\varphi}^t \sqrt{-g} dr  d\theta d\varphi 
=2\pi \lim_{r\to \infty}\int_0^\pi  r^2 \sin \theta A^{(0)}(\theta) V^{(1)}(\theta)  d\theta \ ,
  \end{equation}
	where we use 
	also  Maxwell's equations together with assumed asymptotic behaviour
	of the metric and matter functions.
 
\subsubsection{Numerical results}

We have studied in a systematic the solutions with boundary data
corresponding to the superposition of a $p=1$ magnetic mode and  an
$\ell=2$ electric mode.
Thus we impose as boundary conditions at infinity
\begin{eqnarray}
V^{(0)}(\theta)=c_e \left( \frac{3}{2}\cos^2\theta- \frac{1}{2}\right)\ , \qquad A^{(0)}(\theta)=c_m L \sin^2\theta \ ,
\end{eqnarray}
where $c_e,c_m$ are input parameters.

\begin{figure}[h!]
\begin{center}
\includegraphics[width=0.6\textwidth]{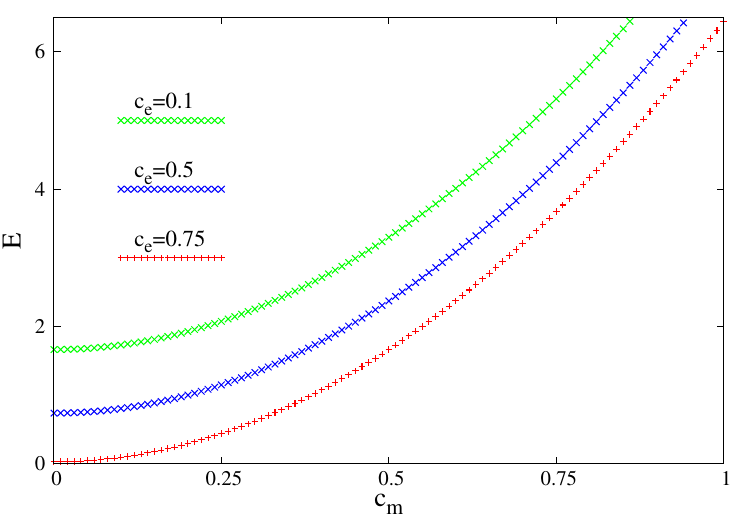}
\includegraphics[width=0.6\textwidth]{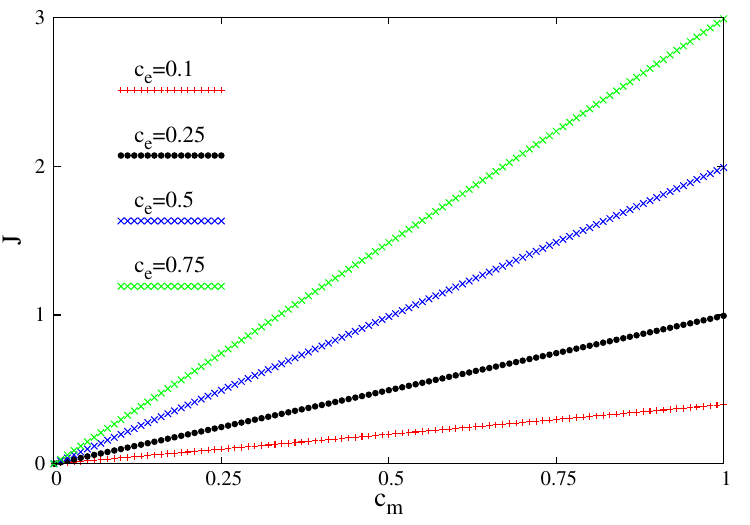}
\caption{Energy (top panel) and total angular momentum (bottom panel) of the gravitating
Maxwell spinning solitons with boundary date given by an $p=1$ magnetic and $\ell=2$ electric multipoles, shown as functions of the ``amplitude" $c_m$
of the magnetic potential at infinity
for several values of  $c_e$ (the ``amplitude" of the electric potential at infinity).
Each point in these plots corresponds to a numerically generated solution. }
\label{perturb}
\end{center}
\end{figure}

In our approach, we set $4\pi G=L=1$ and vary the magnitude of $c_m$
for fixed $c_e$, or vice versa. 
Then the numerical results clearly indicate the existence of everywhere regular, finite energy and angular momentum 
solutions corresponding to $Type~I$
spinning solitons. 
In Fig.~\ref{perturb} the energy and total angular momentum of several families of
solutions
 are exhibited for different values of $c_e,c_m$. Taking $c_e=0$ (or $c_m=0$) these reduce to static, purely magnetic 
(or purely electric) configurations. We remark that although at leading order in a $1/r$ expansion, the Maxwell potential corresponds to the superposition of a $p=1$ magnetic mode and  an $\ell=2$ electric mode only,
 the next order terms ($1/r$ momenta) are already a superposition of all  $(\ell,p) \geq 1$ modes. 
This feature can be anticipated from the perturbative solutions.
 
Also, for all $Type~I$ solutions constructed so far, 
the distribution of the energy and angular momentum
is qualitatively similar to that shown in Figure 1,
with the existence of a single extrema (located at the origin) 
for both mass and angular momentum densities.
On the other hand, we have preliminary results for the existence of $Type~II$
 solitons with boundary data
corresponding to the superposition of a $p=1$ magnetic mode and  an
$\ell=4$ electric mode.
As anticipated by the results in the test field limit, these spacetimes
rotate  locally ($T_\varphi^t\neq 0$) but not globally ($J=0$). 
 A more systematic study of the 
spinning Einstein--Maxwell-$AdS$ solitons, for different boundary data, will be discussed elsewhere.
 
\section{Remarks}
\label{sec_5}
In this letter we have shown that Einstein--Maxwell--$AdS$ theoy admits spinning solitons. 
There are 
 no analogue objects  to these solutions in asymptotically flat spacetime
and
their existence can be traced back to the ``box"-like behaviour of the AdS spacetime.
This fact can be simply understood by the electrostatics-magnetostatics type analysis presented in~\cite{Herdeiro:2015vaa} and further developed here. 
The only mechanism known to yield gravitating solitons with a spin-1 Abelian field
minimally coupled to Einstein's gravity in asymptotically flat spacetime is to consider a Proca, rather than Maxwell, field and take it to be complex, yielding the recently found Proca stars~\cite{Brito:2015pxa}. The latter can also spin, but trivialize in the flat space limit, unlike the solutions discussed here.\footnote{A spin-1 \textit{non-Abelian} field possesses gravitating particle-like solutions which are asymptotically Minkowski~\cite{Bartnik:1988am}.
Intriguingly, these solitons do not allow for spinning generalizations \cite{VanderBij:2001nm}. With $AdS$ asymptotics, however, the state of affairs is different~\cite{Radu:2002rv}.} 

 As a final remark, all solutions of the Einstein--Maxwell--$AdS$ model~\eqref{EMAdS} can be uplifted to eleven dimensional supergravity~\cite{Cremmer:1978km}, yielding the following line element 
\begin{eqnarray}
ds_{11}^2=g_{\mu\nu}^{(4)}dx^\mu dx^\nu
 +4L^2 \sum_{i=1}^4 
 \left[d\mu_i^2+\mu_i^2(d\phi_i+A_\mu dx^\mu)^2 \right] \ ,
\end{eqnarray}
and 4-form field strength
\begin{eqnarray} 
F^{(4)}=\frac{3}{L}\epsilon^{(4)}+4L^2 \sum_{i=1}^4 d\mu_i^2 d\phi_i \star^4 dA \ ,
\end{eqnarray}
where $\epsilon^{(4)}$ is the volume form of the reduced four 4-dimensional space, 
and $\star^4$ denotes Hodge duality in this space
\cite{Chamblin:1999tk}. 
Thus, the solitons here yield new classes of solutions of eleven dimensional supergravity.

\section*{Acknowledgements}
C. H. and E. R. acknowledge funding from the FCT-IF programme. This work was partially supported by  the  H2020-MSCA-RISE-2015 Grant No.  StronGrHEP-690904, and by the CIDMA project UID/MAT/04106/2013. Computations were performed at the Blafis cluster, in Aveiro University.

\bigskip


 \begin{small}
 
 \end{small}

\end{document}